# Understanding Complex Service Systems Through Different Lenses: An Overview

Gerard Briscoe[a,*], Krista Keränen[b], Glenn Parry[c]

[a] Computer Laboratory, University of Cambridge, 15 JJ Thomson Avenue, Cambridge CB3 0FD, UK
[b] Institute for Manufacturing, University of Cambridge, 17 Charles Babbage Road, Cambridge, CB3 0FS, UK
[c] Bristol Business School, University of West England, Coldharbour Lane, Bristol, BS16 1QY, UK

**Summary** The 2011 Grand Challenge in Service conference aimed to explore, analyse and evaluate complex service systems, utilising a case scenario of delivering on improved perception of safety in the London Borough of Sutton, which provided a common context to link the contributions. The key themes that emerged included value co-creation, systems and networks, ICT and complexity, for which we summarise the contributions. Contributions on value co-creation are based mainly on empirical research and provide a variety of insights including the importance of better understanding collaboration within value co-creation. Contributions on the systems perspective, considered to arise from networks of value co-creation, include efforts to understand the implications of the interactions within service systems, as well as their interactions with social systems, to co-create value. Contributions within the technological sphere, providing ever greater connectivity between entities, focus on the creation of new value constellations and new demand being fulfilled through hybrid offerings of physical assets, information and people. Contributions on complexity, arising from the value co-creation networks of technology enabled services systems, focus on the challenges in understanding, managing and analysing these complex service systems. The theory and applications all show the importance of understanding service for the future.

## 1 Introduction

Complex service systems have been identified as a significant area for development in the study of service and led to the first ever conference on the subject, the 2011 Grand Challenge in Service, at University of Cambridge. The conference aimed at furthering the understanding of complex service systems through different lenses. It formed part of the Cambridge Service Alliance annual Service Week and was organised under the Academic leadership of Professor Irene Ng, Professor of Marketing and Service Systems at the University of Warwick. The conference drew global experts from across industry and academia. The design, management and delivery of complex service systems to achieve service excellence and economic viability suggests the need to fully understand the configuration of resource, which includes people, complex equipment, technology and processes. There is therefore a need to understand the theory and practice of complex service systems as well as the value propositions that constitute them. With a dynamic and rapidly changing business environment, organisations are confronted with many challenges as they try to develop their capability for complex service solution provision and deliver on the promise inherent in new business models.

* Corresponding author
E-mail address: gerard.briscoe@cl.cam.ac.uk

Unique to conference is a requirement for participants to apply their expert knowledge to a common case scenario. The case scenario is "Reducing the Fear of Crime in a Community as a Complex Service System: The Case of London Borough of Sutton" [Andreu et al., 2011], in which the Safer Sutton Partnership Service (SSPS) is a complex public sector local authority attempting to deliver an outcome on the perception of safety in a region of South London. A 31 page case scenario synopsis and 291 pages of supporting documentation are provided to all participants to examine and link to their findings. The conference was opened by Warren Shadbolt, the Executive Head of Community Safety and Youth Engagement of London Borough of Sutton, who is ultimately charged with delivery of the complex SSPS provision.

This paper provides an overview of the range of contributions made to the conference and the remainder of the paper is organised as follows. First, a background to the development of complex service systems in service research is presented. This section is then followed by sections on each of the key themes that emerged from the conference and a summary of the contributions to the case scenario. After which conclusions are drawn and possible future research directions are proposed.

## 2 Background

Service Science applies scientific understanding to advance the ability to design, improve, and scale service systems for business and societal purposes [Maglio 2008]. Aiming to integrate elements of business strategy, management sciences research, computer science operations, industrial engineering, social and legal sciences, and others Service Science seeks to encourage innovation in how organisations create value for and with customers and stakeholders that could not be achieved through such disciplines working in isolation [IfM 2008].

The evolution of multidisciplinary service research can be characterised through five time periods [Fisk 1993, IfM 2008, Gummesson 2010]. The first period from 1950 until 1980 is called the *Crawling out* period [Keränen 2011], the initial phase in which service marketing and service operations became distinct from goods marketing and operations. Much of the research and discussion focuses on the question of how services differ from goods. The classic distinctions between services and goods are intangibility, heterogeneity, inseparability (simultaneous production and consumption), customer participation and perishability [Shostack 1977].

The second period from 1980 until 1985 is called the *Scurrying About* period [Fisk 1993, Keränen 2011], during which time a core group of service academics and business practitioners developed [Grönroos 1985, Lovelock 1984, Shostack 1981]. Services research moved beyond the goods and products dyad, though remained mostly conceptual. The literature highlights, for example, the needs for management of the personnel and customers who are involved in the service experience, the physical aspects of the service and the process by which the service is delivered [Booms 1981].

The third period between 1986–1992 is called the *Walking Erect* period [Fisk 1993, Keränen 2011] in which several models describing the process of new service development emerge [Donnelly 1985, Scheuing 1989, Bowers 1986]. Other emerging topics include, for example, issues such as service quality [Grönroos 1983, Parasuraman 1985], the design and management of service production and encounters [Czepiel 1985, Eiglier 1987] and the role of customers, intangibles, and the physical environment in the customer's evaluation of the services [Larsson 1989, Hui 1991].

The fourth period from 1993 until 2000 is called the *Making Tools* period [IfM 2008, Keränen 2011] in which service research was broadened, deepened and sharpened, becoming more quantitative, including measurement, statistics and decision support modelling. There is multidisciplinary research and expanded topic areas include service productivity [Ojasalo 1999], service experiences, service quality and customer satisfaction, connecting operational factors that affect quality to customer loyalty and service orientation, service supply chains, service recovery, technology infusion and service computing [IfM 2008].

The period since the turn of the millennium denotes the fifth period, called the *Creating Language* period [IfM 2008, Keränen 2011], in which new models of service are emerging and the concept of a service system is beginning to take hold, uniting the many perspectives within service science. The field is expanding rapidly with increasing numbers of worldwide researchers and conferences, centres and networks. Initiatives such as the Service Science, Management and Engineering (SSME) aim to strengthen industry-academic-government interactions [Hefley 2008]. Service science has come to include the application of scientific understanding to advance the ability to design, improve and scale service systems for business and societal purposes [Maglio 2008]. The Service-Dominant logic view [Vargo 2004, 2008, Gummesson 2008], service logic view [Grönroos 2008, 2011] and the goods logic view of services are gradually replacing the traditional goods-versus-services dyad. Significant applied service research has been undertaken in manufacturing companies as a result of their move towards both a service-oriented approach and an offer of comprehensive customer solutions [Ojasalo 2008].

The University of Cambridge hosted the first Grand Challenge in Service conference in 2010, officially launching the newly founded Cambridge Service Alliance. The conference featured a week of events and also received sponsorship and support from the Cambridge Service Alliance, Advanced Institute of Management (AIM) Research, the University of Exeter Business School, Rolls Royce, Manchester Business School, SSMEnetUK and the University of Arizona's Eller College of Management. The 2010 Grand Challenge in Service Week brought together leading academics, industrialists and policy makers to address the evolving challenges facing service education, research, practice and policy. One of the important concerns stated that future of service research needs to be a move toward an integrated agenda to better understand how people, processes, and assets interact within complex service systems for the co-creation of value with customers. Based on this, one of the questions brought forwards was how to organise the current research environment for better multi/inter-disciplinary research for greater relevance and impact to industry and society, and how should intangible research be transferred to practise. Hence, the discussions during the Service Week 2010 informed the topics and format of the 2011 conference, including the shared case scenario.

At the 2011 conference leaders in the field created an expert panel of service systems thinkers who provided presentations at the start of each day. The conference then divided into nine sessions from which the following key themes arose; Value Co-Creation and Collaboration, Systems and Networks, Information and Communications Technology (ICT), and Complexity. The contributions and links to the case scenario of the SSPS are considered in the following sections.

## 3 Value Co-Creation and Collaboration

According to Spohrer and Maglio [2010] service is value co-creation – broadly speaking, as useful change that results from communication, planning, or other purposeful and knowledge-intensive interactions between distinct entities and service science is the study of value co-creation. In order to collaborate two or more people, institutions, firms or societies need to associate with each other. During interactions a company has an opportunity to engage with its customers' value creation system and become a co-creator of value [Grönroos 2011]. There can be multiple points of interaction in the system and all the points of customer-company interaction are critical for value creation. All points of interaction between the provider and the customer are opportunities for value creation [Prahalad 2004, Ng 2011].

Most of the conference contributions on the co-creation phenomenon include empirical evidence and several different topics are addressed. One of the most important topics raised was that of value resonance among stakeholders, presented by S Sebhatu, M Johnson, H Gebauer, B Enquist. Value resonance arises when a company, foundation, and customer values align; an important factor when creating meaning and legitimating among actors in a complex service system. As examples IKEA and Starbucks both discuss their success through ensuring such alignment exists in their values based service businesses. It is important that customer' values and company values resonate, drawing not only on customers' values but also the company's culture, leadership, and governance. Companies, achieving value resonance require values based on brand and development of interactive communication based on dialogue amongst all parties.

P Vyas and R Young consider how service providers can display pro-social behaviour, initiating and coordinating a service for the wider community of a city or region. This includes socially responsible design as a means to assist collaborative approaches for community engagement. To create value in many-to-many service contexts a structured approach to mass service creation in networks is necessary. One recommendation is to encourage case research and ethnography where possible, so as to enable better understanding of how customers interact with one another to create value for themselves and others.

A variety of views were put forwards as to what is and what is not value co-creation. A major concern was that drawing too narrow a view would be unhelpful. Equally, if all value is co-created the concept becomes generic to the point of redundancy. Consensus was reached by considering the development of a model consisting of different levels of co-creation. Whilst value is always co-created the degree of co-creation changes dependent upon the focus of the system, which encompasses the levels within the model. This key theme leads us to consider the next area that arose from the conference, that of systems and networks.

## 4 Systems and Networks

Systems theory has garnered increasing attention within service research because of contributions to the understanding of service phenomena [Barile 2010]. A service system is shaped by forces in the wider social system and so an unbounded approach to service system analysis allows us to go beyond the conventional frame of reference and take into account a collection of factors that are inseparably linked [Banathy 1996]. Furthermore, we can conceptualise and analyse the network of interactions that define service within a service system framework.

Contributions that were presented and discussed ranged from fundamental efforts to understand service systems to specialised efforts to understand individual industries, such as the automotive industry. The fundamental contributions aim at understanding wider service systems, their boundaries, the dynamics of their inherent cycles and their interactions with the social systems in which they exist and symbiotically sustain through the emergence of value constellations. Many of the approaches presented aim at encompassing a wider systems perspective, including conceptual, modelling and computational frameworks, with some based in structuration theory [Giddens 1984, Sewell 1992] and others on the theory of fuzzy systems, used to provide an understanding of the structures within service and social systems and their influence on creating competitive value propositions. Ontological models of services were also presented that consider the interrelations of states, events and processes occurring in a wider service system, including the interactions between services and their impact on the surrounding social system. Contributions include efforts to better understand what enables service innovation in the networks of interactions that make up service systems. More reflexive contributions aim at helping develop a better appreciation of how to fully understand the wider service systems. This includes perspectives such as critical realism in which external observation to determine actors and relationships needs to be augmented by data capturing the perceptions of those actors within the system. The possible effect of fragmentation from functional division can prevent actors from seeing the wider system of networked interactions.

The implications of the viable systems approach was presented by S Barile, F Polese, M Saviano and P Di Nauta, particularly with respect to the boundaries used to identify possible governance approaches for establishing sustainable relationships among the actors, sharing knowledge and valorising common resources. The start point is a theoretical discussion on the concept of boundary and its implications to service systems understanding, from a systems theory perspective [von Bertalanffy 1950, Parsons 1971]. Then, a Viable Systems Approach [Golinelli 2000, VV 2011] is adopted to take into account recent service research advances such as Service Science [Maglio 2008] and Service-Dominant logic [Vargo 2008]. Wider and more porous systems boundaries are proposed to better understand governance mechanisms and managerial behaviour, because many organisations today are open to external dynamics and need to interact with many other actors (owners of needed resources) in the pursuit of value co-creation.

Computational modelling and simulation of service systems by integrating models and data of component systems together into bigger and more encompassing models was presented by C Kieliszewski, P Maglio and M Cefkin. The platform takes models of real-world systems, synthesising and integrating them into an interoperating composite system model that policy-makers can use to evaluate alternative scenarios, understanding the difference between current and more idealised perceptions for achieving sustainable change. Construction from data and models that are varied in nature and granularity range from statistical and queuing models to agent-based and social network models. This analysis takes service system analyses beyond individual provider-customer dyads in isolation and instead considers entire value constellations, providing mechanisms for cataloging, describing, connecting, and executing diverse sets of models together.

Understanding service provision increasingly requires a system perspective and framework to better understand the networks of interactions through which services emerge, especially with the increasingly complex nature of technology augmented networks. These themes of complexity and technology are explored next.

# 5 Information and Communications Technology

The services industry is of increasing importance to the global economy, with many technology augmented services delivered through complex networks of service providers. The United Kingdom (UK) government published a paper officially acknowledging the development of electronic services as a crucial factor for the UK economy [DTI 1998]. Advances in technology are facilitating greater connectivity between entities, such that services are hybrid offerings of physical assets, information and people.

While the network aspect in services has been long established, increasing connectivity is leading to ever greater complexity, for which technology (ICT) is often both a cause and sometimes a cure. The myriad of services that are ICT dependent (or at least enabled) continues to grow. Maintaining an awareness of the changing state-of-the-art is paramount because of the ever increasing pace of change. New paradigms such as Cloud Computing[1], are an example of this, with opportunities for the improved delivery of ICT-enabled services.

Contributions were presented and discussed which were aim to help to manage the increasing number of large, complex service systems through technology. These efforts include approaches being developed to understand how the use of technologies can lead to the creation of the some of these complex service networks due in part to the increasing pervasiveness of ICT (e.g. mobile applications and broadband internet). Approaches presented include model-based and process tree based learning, consideration of the efforts to better understand and manage the acquisition of information requirements, trust issues, and aspects of quality of service. Consideration was given to how technologies have significantly stretched the scope for knowledge management generally and digital government specifically, allowing differentiation in the services delivered to provide for a diverse range of users, and how this presents new organisational challenges because of the complexity from increased variety within digital service environments. Contributions around knowledge management were made, as the transfer and utilisation/exploitation of knowledge is seen as important for organisational life and growth. The potential for transfers across and between organisations with increasing efficiency through the adoption of new technologies has the potential to accelerate the diffusion of innovation.

The existing and expected impact of new technology paradigms were considered. For example, J Busquets, J Rodon, J Batista, L Soldevila, T Plana, T Aromir, E Gassiot, P Navarro, X Martín and M Montes presented forms of social networking which will become a central element in the client strategies of customer centric organisations in which services and service innovations are paramount in achieving long term competitive advantages. The case of financial institutions is considered; banks generally are customer centric organisations [Cusumano 2010]. Overall, a better understanding of customer intentions with social media and digital platforms is sought when promoting innovations in complex services to differentiate and de-commoditise offerings, with the banking industry as an exemplar.

Another technology paradigm considered is that of service modularity, which is made possible by ICT platforms, presented by R Rajala, V Tuunainen and H Cassab. This includes the concept of service modularity in the design of ICT-enabled services, and the business models required for them. This work is motivated by the lack of sufficient attention to how business model design is connected with service modularity, even though prior research on product innovation management shows that modular product design and organisational ambidexterity contribute to product innovation capability [Tushman 1996]. The approach adopted involves integrating theories of strategic orientation and organisational learning from the innovation management literature with research on service design, information systems and business models to propose a framework for investigating service modularity for ICT intensive service innovations. This research-in-progress establishes a framework with four elements, considered as building blocks in a modular design of any ICT-enabled services: (1) the service offering, (2) resources, (3) the revenue model, and (4) the relationships with intra- and inter-organisational actors. Consideration is then given to how the framework could be used to benchmark organisational efforts towards modular service design and to meet the management challenges associated with ICT-enabled services.

---

[1] Cloud Computing is a marketing term for technologies that provide computation, software, data access, and storage services that do not require end-user knowledge of the physical location and configuration of the system that delivers the services. A parallel to this concept can be drawn with the electricity grid, wherein end-users consume power without needing to understand the component devices or infrastructure required to provide the service [Marinos 2009].

Cloud Computing, an emerging paradigm, provides a compelling value proposition for organisations to outsource their Information and Communications Technology infrastructure [Briscoe 2009]. However, the transformation of an organisation's or a community's ICT-enabled service systems towards a Cloud-Computing-based service model is a complex process. A study presented by N Su aims to create a conceptual framework to facilitate the understanding of such service systems and their transformation towards the Cloud Computing paradigm, developing a framework to conceptualise Cloud Computing within the broader service ecosystem. Such a service ecosystem consists of diverse stakeholders with different options for acquiring different services. This service ecosystem is dynamic, as an increasing amount and variety of services can be transformed towards Cloud Computing. The more standardised, automated, and commoditised services are, the more likely they are to be transformed into the self-served on-demand utility-model that the Cloud Computing paradigm represents.

The realisation of the ICT-enabled services builds upon increasingly interconnected networks of providers to enable service innovation, which can only be fully understood and realised with an appreciation of their complexity, the next key theme discussed.

## 6 Complexity

Complexity within service systems generally arises from their increasing scale and/or number of components. In is proposed that understanding complex service systems benefits from a systems perspective and a wide knowledge base. To meet this challenge, one approach is to establish a network of professionals from different fields to explore, analyse and evaluate complex service systems from different perspectives; an aim of this conference.

Breaking the boundaries between product and service facilitates the development of new ways of service system thinking. Complex services exist as an integral part of complicated services and separating them is not possible. Regarding the case scenario, the SSPS involves over 30 organisations and is recognised as a complex service. However, complexity can exist in even a relationship between only two parties. Complex systems are constantly adapting and their interactions can have emergent effects upon outcomes, which are sometimes unpredictable. This unpredictability requires understanding if sustainable complex systems are to be developed. Typically too much emphasis is placed on the static nouns of a system, rather than the dynamic verbs which are more significant in facilitating the understanding of complex outcomes.

Based on the research presented during the conference even understanding, analysing and modelling complex service systems can be challenging. Work presented aims at addressing the perceived knowledge gap in understanding how institutions, firms and society operate in a dynamic complex service system. Questions raised during the discussions sought definitions of intelligent complex service systems, and how they might be helpful in achieving the stated aim.

A three step approach to management of complex service systems was presented by G Parry, J Mills, V Purchase. The research was part of a study into the provision of military jets under an availability contract and describes tools developed to pragmatically fulfil the needs of service managers. The approach integrates value analysis, the creation of an image capturing the organisations involved, and an approach to identify and address the complexity challenges. Value analysis requires interviews with multiple stakeholders to capture the breadth of different perspectives of partners in the enterprise. Value perspectives reveal the contracted (implicit) and un-contracted (explicit) needs of the stakeholders. This work highlights that complex service systems place heavy reliance on relational exchange to achieve enterprise goals. The visualisation of the stakeholders utilises an approach called Enterprise Imaging, a development of service blueprinting but which includes multiple back office functions to support a common front office where services are co-produced. Complexity analysis was built upon the premise that complexity was a hindrance to system performance, and that companies were engaged because of their capability in managing specific complex service systems. Empirical analysis further identified six core areas likely to be source elements of complexity; operand resource, operant resource, process, contractual, organisational, and financial. These were presented as a guide to managers as to where complexity might exist, thus enabling them to focus limited resource when undertaking targeted change activity.

A detailed visual representation of complex public sector service provision was presented by J Mills, V Purchase, G Parry. This work further developed the concept of Enterprise Imaging and provides descriptions of how a complex service system may be captured. The image framework creates a central front office area, with a back office above and below it, which represents the two main contracting parties. Individual units are drawn upon the framework, placed appropriately in the front or back office depending upon visibility to the other parties. These are organisational units which undertook activity to co-produce service outcomes. The 'final' image provided is an epistemic object around which discussion and strategy may be built rather than a factual representation of an enterprise. The approach was applied to a number of complex service systems, including fast jet support, UK National Health Service (NHS) psychiatric intensive care provision, and the SSPS case scenario. Each organisation involved stated how the process of creating the image was as valuable as the image itself when understanding and contextualising their respective complex service systems.

J Angelis, G Parry, M Macintyre presented work focussed upon discretion and complexity in customer focused environments. Operations have traditionally focussed on reductive analyses in which transactional processes are open to mass-customisation and standardisation. This work proposes that service complexity created by extensive reasonable customer demand limits the ability to standardise and manage systems through mass-customisation. Beyond mass-customisation it is proposed that management is achieved via discretion, which is difficult, if not impossible to codify, so operations are necessarily managed via framework principles, and these must be embedded in culture. This result creates numerous challenges for managers because while providing a source of sustainable competitive advantage it is difficult to replicate and culture and embedded frameworks are difficult to scale across firms. The conclusion proposed that service complexity created by extensive 'reasonable' customer demand, limits the ability to standardise and manage systems through mass-customisation.

The first steps have been taken in understanding, managing and analysing the complexity that can emerge in service systems. The range of contributions shows that the challenge affects many areas of research, although often in different ways.

The opportunity to address the challenge collectively under a common theme presents a way forward.

## 7 Case Scenario

Warren Shadbolt, the Executive Head of Community Safety and Youth Engagement of the London Borough of Sutton, describes the challenge of complex service provision as a public sector provider. The London Borough of Sutton is a globally typical middle class affluent suburb which could be found in any large city. Whilst overall described as affluent, areas of the North of the borough would be described as deprived based on many different public metrics such as teenage pregnancy, though not necessarily crime. The borough has many educational establishments and daily is a net importer of about three thousand people. Due to its overall affluent nature Sutton borough is not eligible for any special government funds to deal with specific issues. However, crime is described as relative and with over fifteen thousand crimes reported a year, around thirty per day, this could have life changing impacts on many people in the borough and is therefore a social problem of concern.

The level of crime was identified as the top priority by residents in an independent annual Ipsos MORI[2] survey, which has been funded for over twenty two years by the Borough council. Though during the most survey, fifty percent of residents had no concerns, which provides evidence of the success of the SSPS initiative. The desired outcome is an improvement in the perceived safety of the borough, which creates some challenges. Gatherings of groups of youths is perceived as threatening, but is not actual criminal activity. This understanding highlights the difference between perceived and actual crime, both of which require attention. SSPS is primarily a Police and local Council initiative, with two hundred and fifty staff working in a holistic blended organisation framework, but which consumes only half a percent of the total revenue budget. This council is believed to be the only one working in this way to co-produce a service. Mr Shadbolt identifies value co-production as a key differentiator in improving the experience of living in the borough, as measured by residents in the survey over the past five years relative to other boroughs. This key differentiator is described as working with the community to define solutions. Real benefits are gained through co-location of service providers, data sharing and objective sharing. Partnerships are described as built on trust, and so continuity of relationships between parties is important. The changes achieved were hard won, with many old guards resistant to the new ways of working. Empowerment is held as a key driver, though it was acknowledged that this was challenging to realise as the public sector has many systems to dampen spending and reduce risk, which can limit individual discretion.

---

[2] Ipsos MORI is the second largest market research organisation in the United Kingdom, formed by a merger of Ipsos UK and MORI, two of the Britain's leading survey companies. Ipsos MORI conduct surveys for a wide range of major organisations as well as other market research agencies.

Contributions to the case scenario from the key themes aim, directly or indirectly, to address the underlying factors in the complex service system that the SSPS represents, and are summarised in Figure 1.

One of the most ambitious contributions was from a group of doctoral students who viewed the case scenario as an opportunity to offer a unique perspective which they felt would be less subject to experience bias of more senior peers. They endeavoured to present a collection of fresh approaches that together would provide a more complete perspective. Each approach ultimately offered a slightly differing interpretation of the case scenario, but all followed a service science theme, ranging from knowledge management to social theory. One approach was centred in service ecology and proposed that perceptions of fear is the fundamental problem, and so suggested solutions based upon the empowerment of individuals within local communities while simultaneously identifying individuals who might have been overlooked for additional attention from the SPSS.

Another contribution took a safety in the community perspective, analysing the risk and safety perceptions of the complex service system using multidimensional scales applied to the citizens. In addition, the work considers the perception of other stakeholders in the community. A research methodology was proposed to analyse perceived safety within a complex service system, with both intra- and inter-variable approaches. Specifically, they consider the use of an index which integrates the multidimensional features of residents' perceived safety, and their interactions with other stakeholders. Furthermore, a research model was suggested that analyses the relationships of perceived safety and other social variables (i.e., satisfaction in the community, social cohesion, quality of life, social participation for value co-creation, etc.). This dual methodology provides important insights into how the fear of crime, as part of a multidimensional perception of safety in the local community, can be subjectively assessed by different audiences, and offers the possibility of making objective comparisons, which would be useful for future planning.

Another contribution adopts a Service-Dominant logic approach, identifying the borough as a service system, to explore residents' perceptions which emerge from their lived experience. Using a customer experience modelling framework residents interactions with other actors are identified, as well as factors that enable or inhibit those interactions. The actual causes of the fear of crime and its changing levels over time could be determined in each of the different council ward areas. Through better understanding the categorisation of these negative emotions and how they change over time, the residents and governing bodies can work together more effectively and tackle the identified experience inhibitors. Adopting a value co-creation approach for the future, governing bodies could regularly update their understanding of residents' experiences and work with them in preventing any future rises in fear within communities.

Consensus within the different themes centred around understanding the psychological aspects of fear, community dynamics, emergent properties and the connection between these. This consensus was exemplified by the individual contributions mentioned which also showed the multi-disciplinary[3] nature of the work, a result of the challenge of viewing the case scenario through different disciplines. This work will hopefully encourage future inter-disciplinary[4] efforts in understanding this and other complex services systems.

---

[3] Multi-disciplinary refers to knowledge associated with more than one existing academic disciplines, such that a multi-disciplinary community can decompose a problem into nearly separable sub-parts, and then addressed via the distributed knowledge in the community.

[4] Inter-disciplinary refers to new knowledge extensions that exist between or beyond existing academic disciplines., such that an inter-disciplinary community is are engaged in creating and applying new knowledge becomes a primary sub-goal of addressing the common problem.

| |
|---|
| **Value Co-Creation and Collaboration** |
| Suggestions included that both the community and the SSPS committee should focus more on value co-creation and co-production, rather than contractual obligations. |
| Considering renewal and values resonance, it was suggested that an involvement of the community (customer) in the production of the service (safety) is a more effective and efficient instrument of value co-creation. |
| The role of design can also help to co-resolve concerns about anti-social behaviour, by seeking to promote greater collaboration and mutual understanding between the stakeholders (police, council staff, residents and youths) to realise an improved level of social wellbeing through co-design. |
| There is a need to identify service creators and customers and discuss the boundaries between the two, because service systems are established to create services with different parties playing an active part in service creation. Customers could become service providers to others. |
| **Systems and Networks** |
| Contributions ranged from considering communicative barriers that might exist from functional divisional within the SSPS to possible dialogues that could be fostered among the actors to encourage constructive service innovation. |
| Given that the key issue is the fear of crime, which is essentially psychological in nature; it could be investigated with a coherent interpretation scheme from a viable systems information variety model [Barile 2011], such that the interpretation schemes could determine the perceived relevance of significant information flows within the system. |
| An approach suggested was that the governance actors within the London Borough of Sutton (Municipality, Safer Neighborhood Inspectors, Ward Councilor, Resident Association Representative, Street Scene Manager, Council Manager) ought to adopt a citizens' viability perspective and so direct governance actions in a sustainable way, investing resources to monitor context conditions that help determine the fear of crime. |
| **Information and Communications Technology** |
| Contributions ranged from considering the role of ICT in achieving the sustainable development of local communities generally, to the potential of creating online communities that can reshape the relationship between different stakeholders of local communities. |
| A possible issue identified was that of trust with the council's ability to carry on tackling crime even though the statistics show that the local authority has been performing well in this area over the few last years. |
| A potential approach considered was the realistic application of ICT-enabled social networking for service innovation, exploring aspects such as culture and innovative contexts for extrinsic motivations relative to social obligations in helping to manage crime and the fear of crime. |
| **Complexity** |
| A significant recommendation was the application and integration of three different approaches (Enterprise Imaging, Value definition, Complexity Analysis) to provide a holistic view of the challenges that need to addressed. For example, reducing the fear of crime can be regarded as an emergent consequence of a complex service system, where the outcome (value) of the service is feeling safe and secure. |
| The three stage model for developing a Product Service System (PSS) could be applied to the case scenario to better understand the complexity of the interactions that gives rise to fear, and therefore how to offer a service that can mitigate and manage this fear, even when emerging non-linearly. |

*Figure 1: Summary of Contributions to the Case Scenario*

## 8 Conclusions

The development of service science and the phenomenon of complex service systems have been considered. The conference contributions of both industry and academia are centred around four key themes: value co-creation and collaboration; systems and networks; information and communications technology; and complexity.

Value co-creation can be seen from many different perspectives, but how broadly or narrowly remains unclear despite constructive contributions suggesting a model consisting of different levels of co-creation. The challenge comes less from the word itself and more from its modifier which indicates collaboration. Observing co-creation through various lenses, rather than from a single perspective has the potential to provide much broader insights into the diverse phenomena encompassed. The contributions of the conference provide a variety of insights into value co-creation, adding to the growing body of research that has emerged in recent years in this area.

Service provision increasingly requires a systems perspective and framework to better understand the networks of interactions through which services emerge, especially with the increasingly complex nature of technology augmented networks. Contributions towards this include efforts to understand the implications of the interactions within service systems, as well as their interactions with social systems.

The realisation of ICT-enabled services arises from the ever greater connectivity allowed between entities, with new demand being fulfilled through hybrid offerings of physical assets, information and people. Contributions of understanding within this sphere highlight the dual nature of complex service system, as they are often both the cause of the inherent complexity and the key to managing complexity.

Contributions on complexity arising from networks of service systems takes the first steps in understanding, managing and analysing the complexity that can emerge. Work is presented with the aim of addressing the perceived knowledge gap in understanding how institutions, firms and society operate in a dynamic complex service system. The range of contributions shows that the challenge affects many areas of research, although often in different ways. The opportunity to address the challenge collectively under a common theme presents a way forward.

Connecting the diverse research within complex service systems is a common case scenario, in which the London Borough of Sutton struggles with the difference between perceived and actual levels of crime within the community. Exploring, analysing and evaluating this shared complex service system from respective disciplines provides a basis for multi-disciplinary collaboration. Despite the differing epistemologies of the contributor's consensus emerged around understanding the psychological aspects of fear, community dynamics, emergent behaviour from complexity and the connections between these aspects. Conclusions identify that collaboration leading to more effective value co-creation is critical when considering the management of complex service systems. The conference and its format show great potential for those considering how to approach real world examples of complex service systems through multi-disciplinary collaboration.

The 2011 Grand Challenge in Service conference provided an opportunity to span knowledge boundaries and provide broader perspectives into service science. A wide range of research was presented, addressing how to understand critical concepts of complex service systems, with richness provided by the diversity of participants' epistemologies. This diversity also shows the multi-disciplinary nature of the contributions. A self-reflexive conclusion suggests seeking more effective value co-creation in research through inter-disciplinary efforts to better understand complex services systems. The conference is therefore a step in the direction for future researching into complex service systems, and also suggests a novel approach to supporting multi- and inter-disciplinary research efforts into complex systems.

Selected contributions were invited to submit extended versions of their papers to this special issue of the European Marketing Journal entitled 'Theoretical and Practice Perspectives in Complex Service Systems'.

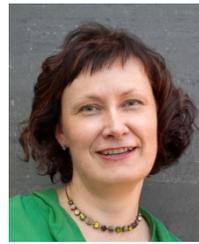

**KRISTA KERÄNEN** holds a Bachelor degrees in Hospitality Management and Business Management, and a Master's degree in Hospitality Management. She is currently doing her PhD at the University of Cambridge, UK, and managing a research project called CoCo. She has a long background in service business and entrepreneurship. She has also been a senior lecturer since 2001, and as a development manager since 2008,, at Laurea University of Applied Sciences, Leppävaara Unit, Finland.

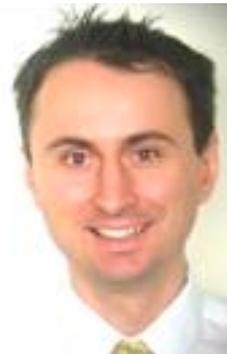

**GLENN PARRY** is Associate Professor in Strategy and Operations Management at Bristol Business School, University of West England. His work focuses on the challenge of delivering customer value and management of complexity. His research has been developed with and applied by leading companies from the aerospace, automotive, music and health sectors. He has published in numerous international journals and is editor of the books 'Service: Design and Delivery', 'Complex Engineering Service Systems: Concepts and Research' and 'Build To Order: The Road to the 5 day Car'.

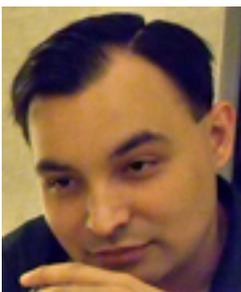

**GERARD BRISCOE** is a Research Associate, Systems Research Group, Computer Laboratory, University of Cambridge. Before this, he was a Postdoctoral Researcher, Department of Media and Communications, London School of Economics and Political Science. He received his PhD in Electrical Engineering from Imperial College London (ICL). He worked as a Research Fellow at the MIT Media Lab Europe, after completing his B/MEng in Computing also from ICL. His research interests include sustainable computing, cloud computing, social media and natural computing.